\documentclass{PoSaa}

\usepackage{subfigure}

\title{A new mechanism for generating broadband pulsar-like polarization}

\ShortTitle{A new mechanism for generating broadband pulsar-like polarization}

\author{Houshang Ardavan,$^1$ Arzhang Ardavan,$^2$ Joseph Fasel,$^3$ John Middleditch,$^3$ Mario~Perez,$^3$ Andrea Schmidt,$^3$ John Singleton$^3$\\
       $^1$Institute of Astronomy, University of Cambridge - UK\\ 
       $^2$Clarendon Laboratory, Oxford University - UK\\
       $^3$Los Alamos National Laboratory - USA}

\abstract{Observational data imply the presence of superluminal electric currents in pulsar magnetospheres.  Such sources are not inconsistent with special relativity; they have already been created in the laboratory.  Here we describe the distinctive features of the radiation beam that is generated by a rotating superluminal source and show that
\begin{enumerate}
\item[(i)] it consists of subbeams that are narrower the farther the observer is from the source: subbeams whose intensities decay as $1/R$ instead of $1/R^2$ with distance ($R$),
\item[(ii)] the fields of its subbeams are characterized by three concurrent polarization modes: two modes that are `orthogonal' and a third mode whose position angle swings across the subbeam bridging those of the other two,
\item[(iii)] its overall beam consists of an incoherent superposition of such coherent subbeams and has an intensity profile that reflects the azimuthal distribution of the contributing part of the source (the part of the source that approaches the observer with the speed of light and zero acceleration),
\item[(iv)] its spectrum (the superluminal counterpart of synchrotron spectrum) is broader than that of any other known emission and entails oscillations whose spacings and amplitudes respectively increase and decrease algebraically with increasing frequency, and
\item[(v)] the degree of its mean polarization and the fraction of its linear polarization both increase with frequency beyond the frequency for which the observer falls within the Fresnel zone. 
\end{enumerate}
We also compare these features with those of the radiation received from the Crab pulsar.}

\FullConference{}

\begin{document}

\section{Introduction}
\label{sec:1}
The rigid rotation of the overall distribution pattern of the pulsar emission reflects a radiation field ${\bf E}$ whose cylindrical components depend on the cylindrical coordinates $(r,\varphi,z)$ and time $t$ as
\begin{equation}
E_{r,\varphi,z}(r,\varphi,z;t)=E_{r,\varphi,z}(r,\varphi-\omega t,z),
\label{eq:1}
\end{equation}
where $\omega$ is the rotation frequency of the pulsar.  Such a field can only arise from an electric current whose density ${\bf j}$ likewise depends on the azimuthal angle $\varphi$ in the combination $\varphi-\omega t$ only:
\begin{equation}
j_{r,\varphi,z}(r,\varphi,z;t)=j_{r,\varphi,z}(r,\varphi-\omega t,z)
\label{eq:2}
\end{equation}
(see appendixes A and B of~\cite{r1}).  This property of the emitting current follows not only from the observational data, but also from the numerical models of the magnetospheric structure of an oblique rotator; it is found that any time-dependent structures in such models rapidly approach a steady state in the corotating frame~\cite{r2}. 

Unless there is no plasma outside the light cylinder, therefore, the macroscopic distiribution of the emitting current in the magnetosphere of a pulsar should have a superluminally rotating pattern in $r>c/\omega$ (where $r$ is the radial distance from the axis of rotation and $c$ is the speed of light {\it in vacuo}).  Such a source is not inconsistent with special relativity.  The superluminally moving pattern is created by the coordinated motion of aggregates of subluminally moving particles~\cite{r3}.  It has been experimentally verified, on the other hand, that such moving charged patterns act as sources of radiation in precisely the same way as any other moving sources of electromagnetic fields~\cite{r4,r5,r6,r7}.

To illustrate the distinctive features of the emission from a superluminal source, we consider a polarization current whose distribution pattern rotates and oscillates at the same time: ${\bf j}=\partial{\bf P}/\partial t$ for which
\begin{equation}
P_{r,\varphi,z}(r,\varphi,z,t)=
s_{r,\varphi,z}(r,z)\cos(m{\hat\varphi})\cos(\Omega t),
\qquad -\pi<{\hat\varphi}\le\pi,
\label{eq:3}
\end{equation}
and ${\hat\varphi}\equiv\varphi-\omega t$.
Here, $P_{r,\varphi,z}$ are the components of the polarization ${\bf P}$
in a cylindrical coordinate system based on the axis of rotation,
${\bf s}(r,z)$ is an arbitrary vector function with a finite 
support in $r>c/\omega$,
$m$ is a positive integer, and $\Omega$ is an angular 
frequency whose value differs from an integral multiple of 
the rotation frequency $\omega$. 
This is a generic source:
one can construct any distribution
with a uniformly rotating pattern,
$P_{r,\varphi,z}(r,{\hat\varphi},z)$,
by the superposition over $m$
of terms of the form $s_{r,\varphi,z}(r,z,m)\cos(m{\hat\varphi})$.
It also corresponds to laboratory-based sources
that have been used in experimental demonstrations
of some of the phenomena described below~\cite{r5,r6}. 

The results reported here are derived from the retarded solution of Maxwell's equations for the above current distribution (see~\cite{r1,r8,r9,r10,r11,r12}).

\section{The field generated by a constituent volume element of the source} 
\label{sec:2}

\begin{figure}
\centering
\subfigure{\includegraphics[height=3cm]{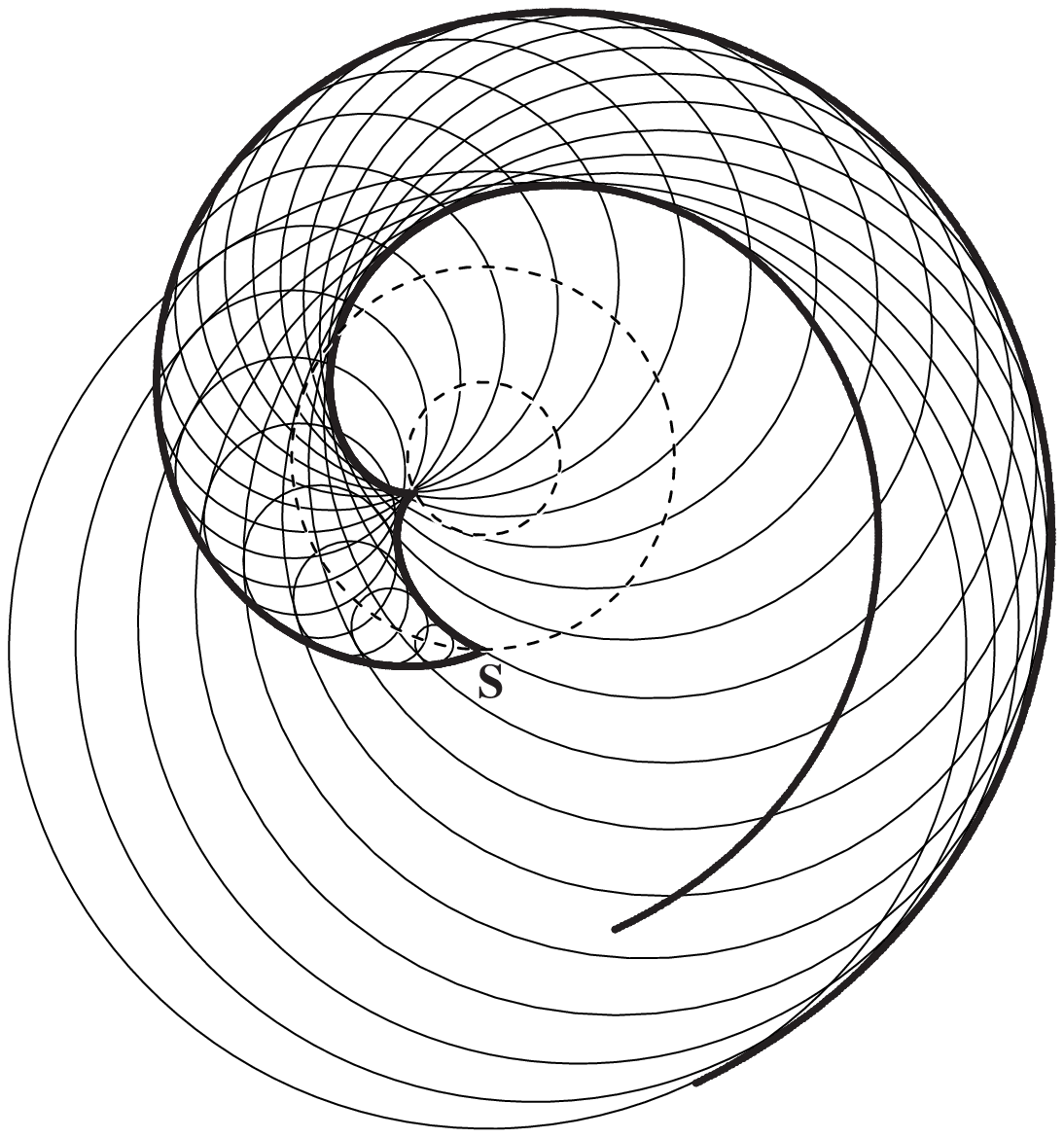}}
\subfigure{\includegraphics[height=3cm]{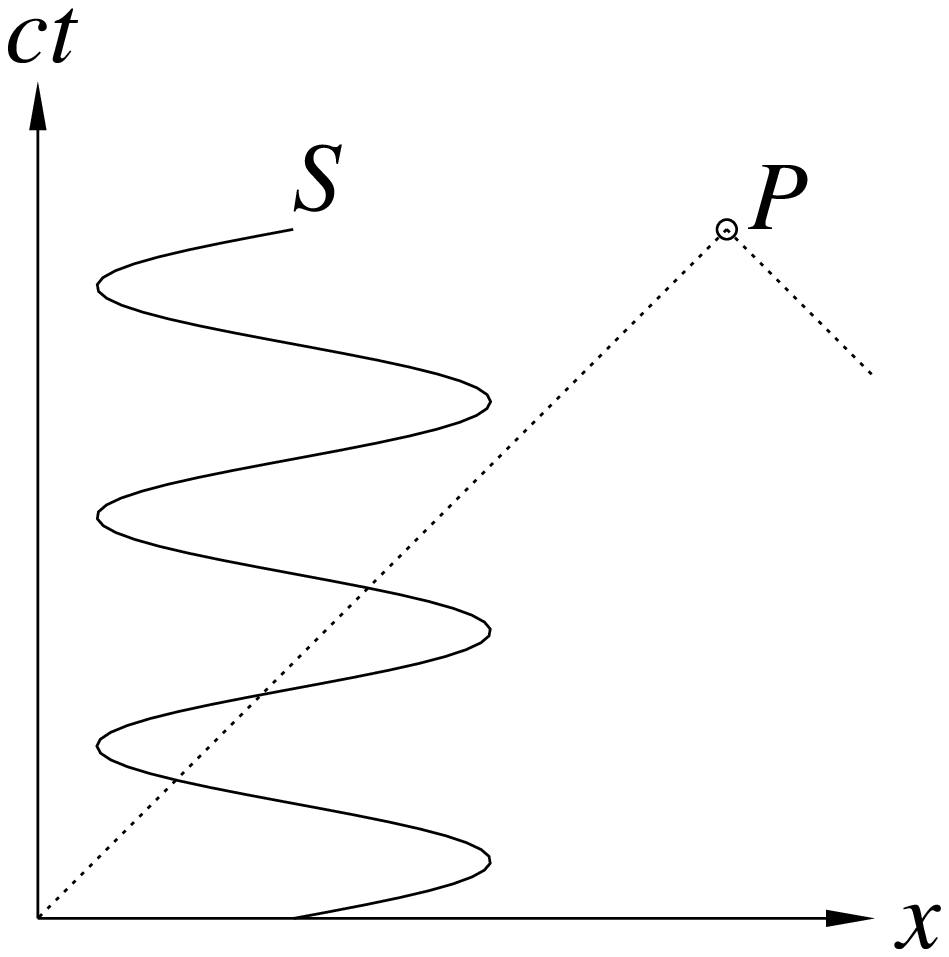}}
\subfigure{\includegraphics[height=3cm]{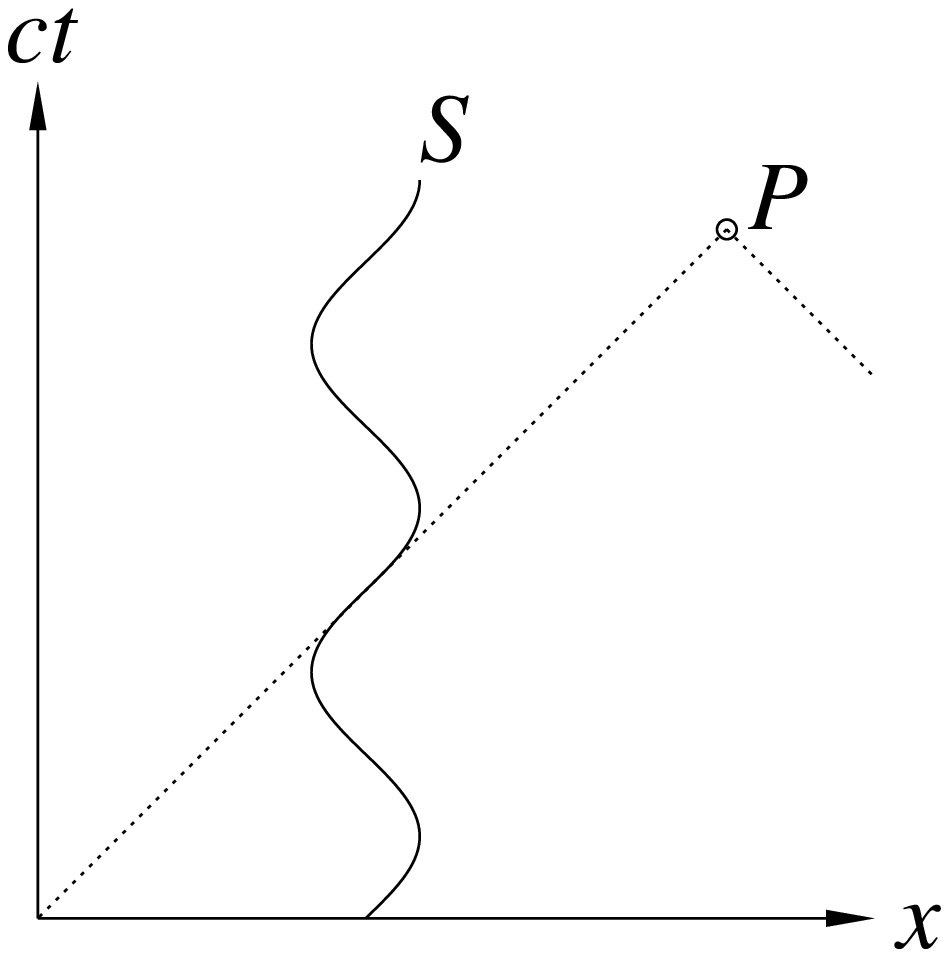}}
\caption{Left: envelope of the spherical wave fronts 
emanating from a superluminally moving source 
element (S) in circular motion.  The heavy curves 
show the cross section of the envelope with the 
plane of the orbit of the source.  The larger 
of the two dotted circles designates the orbit 
and the smaller the light cylinder.  Middle and right: the spacetime diagrams showing the intersection of the trajectory of the source point $S$ with the past light cone of the observation point $P$ when $P$ lies inside, and on the cusp of, the envelope of wave fronts, respectively.}
\label{fig1}
\end{figure}

A superluminal source is necessarily volume-distributed~\cite{r3}.  However, its field can be built up from the superposition of the fields of its constituent volume elements which are point-like.  Figure~\ref{fig1} shows that the waves generated by a constituent volume element of a rotating superluminal source possess a cusped envelope and that, inside the envelope, {\it three} wave fronts pass through any given observation point simultaneously.  This reflects the fact that the field inside the envelope receives simultaneous contributions from three distinct values of the retarded time (see Fig.\ \ref{fig1}).  On the cusp of the envelope, where the spacetime trajectory of the source is tangent to the past light cone of the observer (Fig.\ \ref{fig1}), all three contributions toward the value of the field coalesce~\cite{r8,r9,r10}.  

On this cusp (caustic), the source approaches the observer with the speed of light and zero acceleration at the retarded time, i.e.\ ${\rm d}R(t)/{\rm d}t=-c$ and ${\rm d}^2R(t)/{\rm d}t^2=0$,
where $R(t)\equiv\vert{\bf x}(t)-{\bf x}_P\vert$ is the distance between the source point ${\bf x}(t)$ and the observation point ${\bf x}_P$. As a result, the interval of emission time for the signal carried by the cusp is much longer than the interval of its reception time~\cite{r9}.

\begin{figure} 
\centering
\includegraphics[height=4cm]{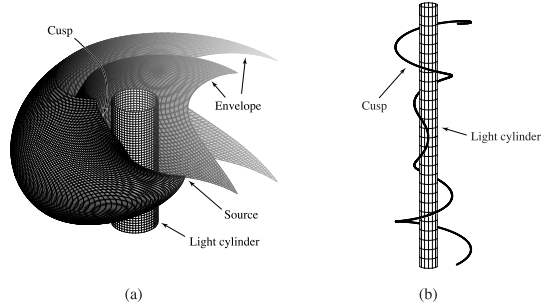}
\caption{Three dimensional views of the envelope (a) and its cusp (b).}
\label{fig2}
\end{figure}

A three-dimensional view of the envelope of wave fronts and its cusp is shown in Fig.\ \ref{fig2}.  The two sheets of the envelope, and the cusp along which these two sheets meet tangentially, spiral outward into the far zone.  In the far zone, the cusp lies on the double cone $\theta_P=\arcsin[c/(r\omega)]$, $\theta_P=\pi-\arcsin[c/(r\omega)]$, where $(R_P, \theta_P, \varphi_P)$ denote the spherical polar coordinates of the observation point $P$.  Thus, a stationary observer in the polar interval $\arcsin[c/(r\omega)]\le\theta_P\le\pi-\arcsin[c/(r\omega)]$ receives recurring pulses as the envelope rotates past him/her~\cite{r11}.

\begin{figure}
\centering
\includegraphics[height=7cm]{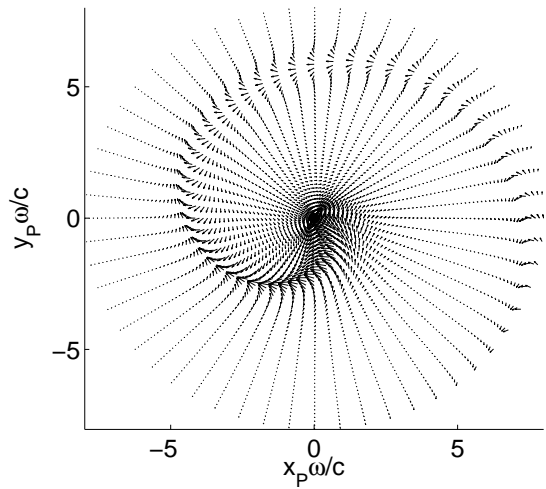}
\caption{Polarization position angles and field strengths on the cone $\theta_P=\pi/12$ outside the envelope for a source with $r\omega=2$.}
\label{fig3}
\end{figure}

Figure~\ref{fig3} shows the radiation field generated by the rotating source element $S$ on a cone close to the cusp, just outside the envelope.  Not only does the spiralling cusp embody a recurring pulse, but the plane of polarization of the radiation swings across the pulse~\cite{r12}.   

\begin{figure}
\centering
\includegraphics[height=5cm]{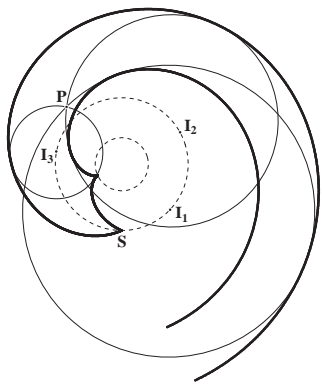}
\caption{An observer $P$ inside the envelope detects three images ($I_1$, $I_2$, $I_3$) of the source $S$ simultaneously.}
\label{fig4}
\end{figure}

As a cosequence of the multivaluedness of the retarded time, three images of the source are observeable inside the envelope at any given observation time (Fig.\ \ref{fig4}).  The waves that were emitted when the source was at the retarded positions $I_1$, $I_2$ and $I_3$ in Fig.\ \ref{fig4} are all received simultaneously at the observation point $P$.  These images are detected as distinct components of the radiation~\cite{r13}. 

\begin{figure}
\centering
\subfigure{\includegraphics[width=7cm]{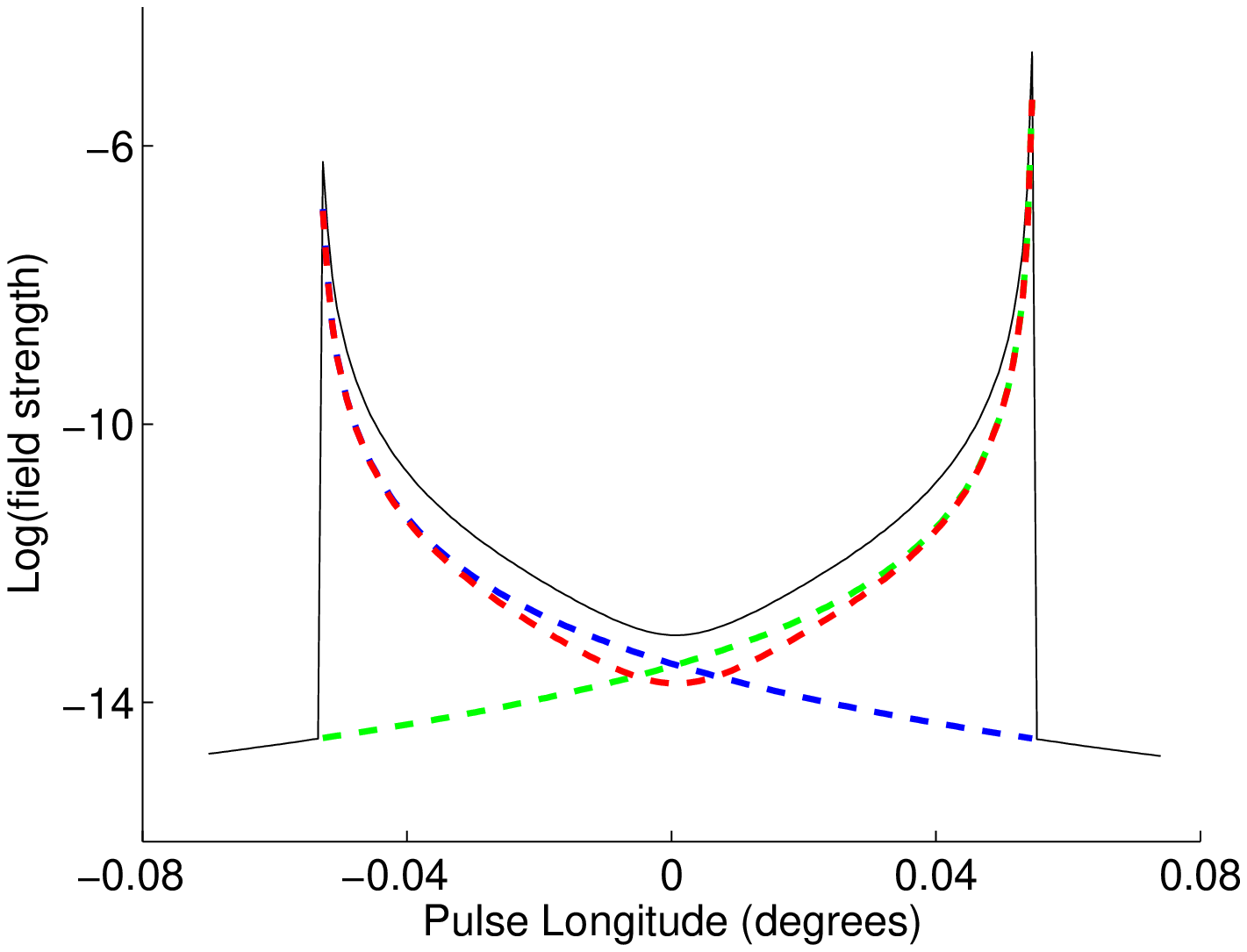}}
\subfigure{\includegraphics[width=7cm]{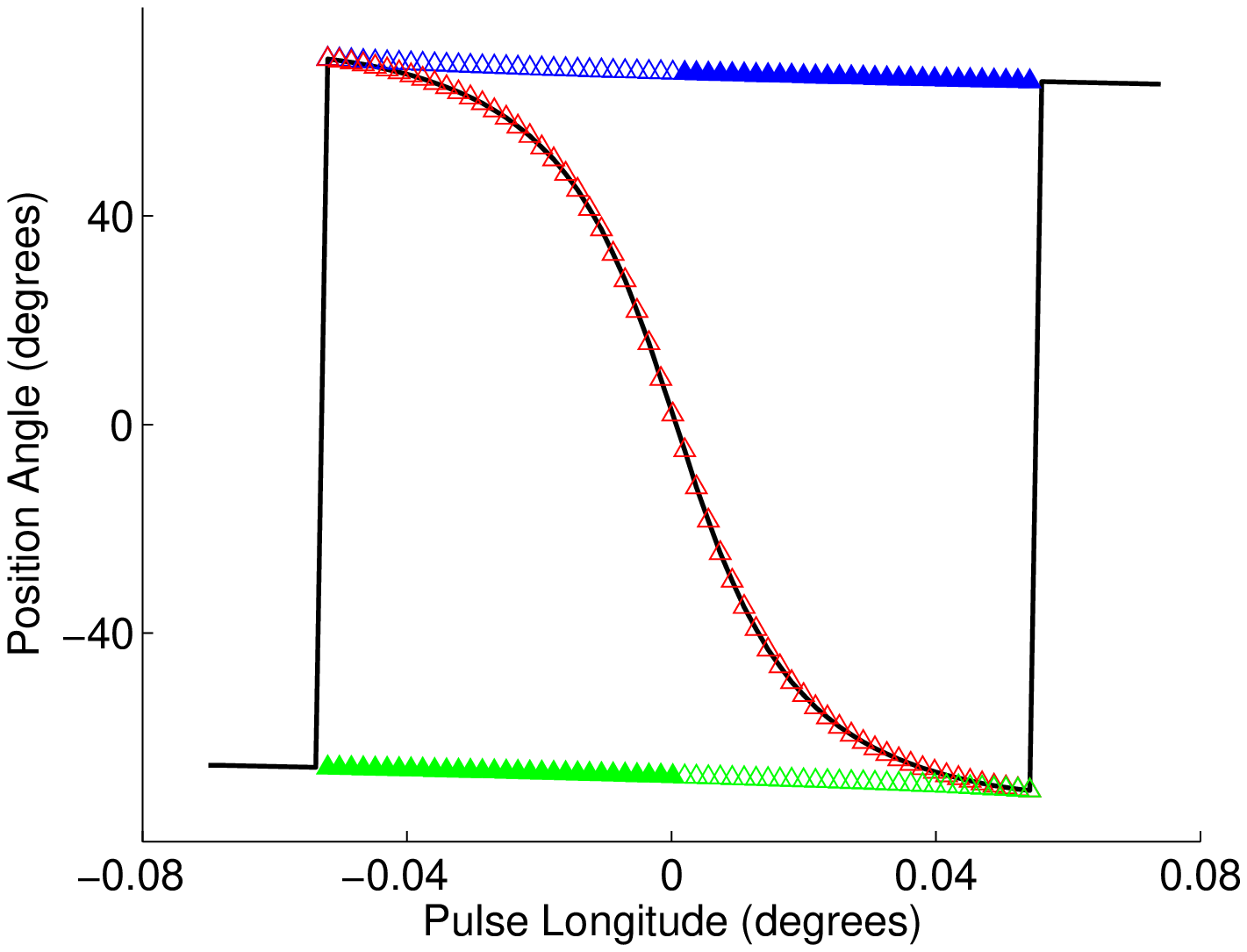}}
\caption{Left: the relative strengths of the three radiation modes as observed near the cusp on a sphere of large radius.  The total field strength (black) and strengths of the underlying contributions from the three images of the source (green, red, blue) are shown for a source with $r\omega=1.1$ and an observation point that sweeps a small arc of the circle $R_P\omega/c=10^{10}$, $\theta_P=\pi/2.7$, crossing the envelope near the cusp. Right: the corresponding position angles of the contributions from the three retarded times (green, red, blue) are shown relative to one another and to that of the total field (black); the position angles of the dominant contributions are shown with open triangles, and those of the weakest contributions with filled triangles.}
\label{fig5}
\end{figure}

Field strengths and polarization position angles of the three images (radiation modes) close to the cusp are shown in Fig.\ \ref{fig5}.  Two modes dominate everywhere except in the middle of the pulse.  Moreover, the position angles of two of the modes are `orthogonal' and that of the third swings across the pulse bridging the other two.  The constructive interference of the emitted waves on the envelope (where two of the contributing retarded times coalesce) and on its cusp (where all three of the contributing retarded times coalesce) gives rise to the divergence of the field of a point-like source on these loci~\cite{r8}.  Here we plot the spatial distribution of the field excising the narrow regions in which the magnitude of the field exceeds a certain threshold~\cite{r12}.

\section{The nondiffracting subbeams comprising the overall beam}
\label{sec:3}

\begin{figure}
\centering
\includegraphics[height=4cm]{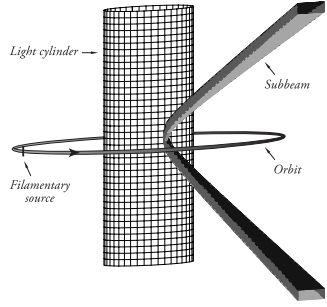}
\caption{Schematic illustration of the light cylinder, the filamentary part of the source that approaches the observation point with the speed of light and zero acceleration at the retarded time, the orbit $r=c/(\omega\sin\theta_P)$ of this filamentary source, and the subbeam formed by the bundle of cusps that emanate from the constituent volume elements of this filament.}
\label{fig6}
\end{figure}

The dominant contribution towards the field of an extended source comes from a thin filamentary part of the source that approaches the observer, along the radiation direction, with the speed of light and zero acceleration at the retarded time~\cite{r11}.  For an observation point $P$ in the far zone with the coordinates $(R_P,\theta_P,\varphi_P)$, this filament is located at $r=(c/\omega)\csc\theta_P$, $\varphi=\varphi_P+3\pi/2$ and is essentially parallel to the rotation axis (Fig.~\ref{fig6}).  The collection of cusps of the envelopes of wave fronts that emanate from various volume elements of the contributing filament form a subbeam whose polar width is nondiffracting: the linear dimension of this bundle of cusps in the direction parallel to the rotation axis remains the same at all distances from the source, so that the polar angle $\delta\theta_P$ subtended by the subbeam decreases as ${R_P}^{-1}$ with increasing $R_P$ (see Fig.~\ref{fig6}).  

In that it consists of caustics and so is constantly dispersed and reconstructed out of other waves, the subbeam in question radically differs from a conventional radiation beam~\cite{r8}.  The narrowing of its polar width (as ${R_P}^{-1}$) is accompanied by a more slowly diminishing intensity (an intensity that diminishes as ${R_P}^{-1}$ instead of ${R_P}^{-2}$ with distance), so that the flux of energy across its cross sectional area remains the same for all $R_P$~\cite{r11}.  This slower rate of decay of the emission from a superluminally rotating source has been confirmed experimentally~\cite{r5,r6}.

The contributing part of an extended source (the filament that approaches the observation point with the speed of light and zero acceleration) changes as the source rotates (see Fig.\ \ref{fig6}).  In the case of a turbulent plasma with a superluminally rotating macroscopic distribution, therefore, the overall beam within which the narrow, nonspherically decaying radiation is detectable would consist of an incoherent superposition of coherent, nondiffracting subbeams with widely differing amplitudes and phases (similar to the train of giant pulses received from the Crab pulsar~\cite{r14}).  

The overall beam occupies a solid angle whose polar and azimuthal extents are independent of the distance $R_P$.  It is detectable within the polar interval $\arccos[(r_<\omega/c)^{-1}]\le\vert\theta_P-\pi/2\vert\le\arccos[(r_>\omega/c)^{-1}]$, where $[r_<,r_>]$ denotes the radial extent of the superluminal part of the source. The azimuthal profile of this overall beam reflects the distribution of the source density around the cylinder $r=c/(\omega\sin\theta_P)$, from which the dominant contribution to the radiation arises~\cite{r11}. 

\section{Superluminal counterpart of synchrotron spectrum}
\label{sec:4}

\begin{figure}
\centering
\subfigure{\includegraphics[width=6cm]{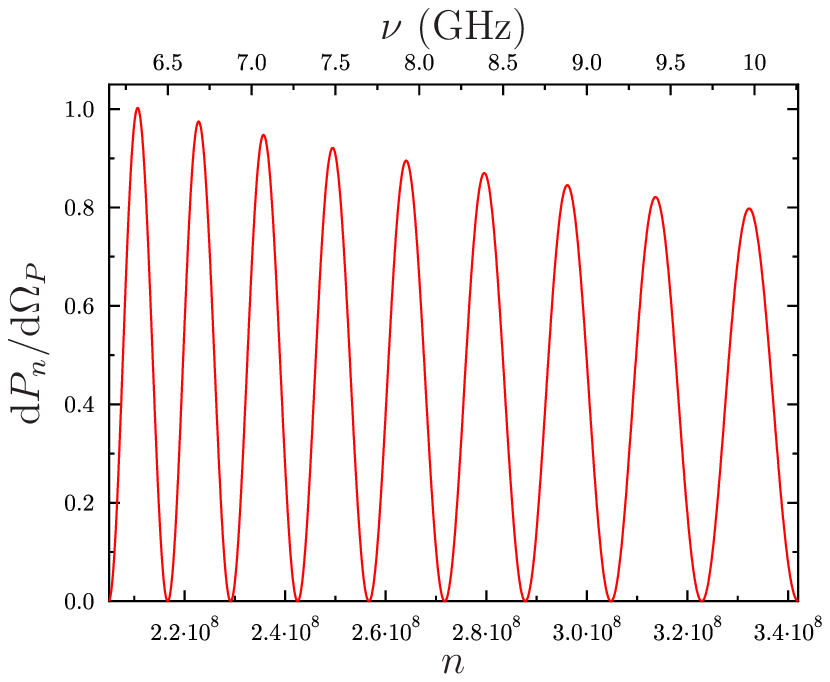}}
\subfigure{\includegraphics[width=6cm]{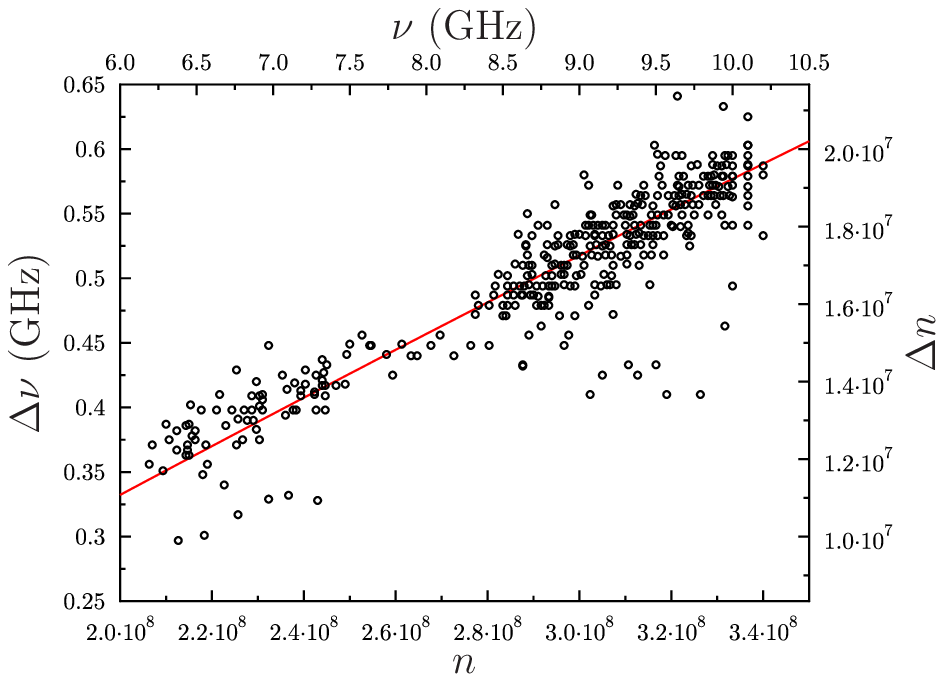}}
\caption{The predicted oscillations of the spectrum of the emission for 
$\omega/(2\pi)=30$ Hz and $\Omega/\omega=1.9\times10^4$ (left) have the same spacing as those of the emission bands in the observed spectrum of the Crab pulsar~\cite{r15} (right).}
\label{fig7}
\end{figure}

\begin{figure}
\centering
\includegraphics[height=6cm]{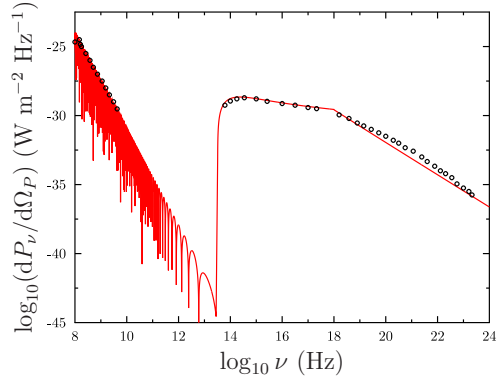}
\caption{The data points show the continuum spectrum of the Crab pulsar~\cite{r16}.  In the model, the recovery of intensity at the ultraviolet peak is caused by resonance with $m\omega/(2\pi)\simeq3$ THz.  The spectral break at $\sim10^{18}$ Hz reflects a transition across the boundary of the Fresnel zone (see~\cite{r1}).}
\label{fig8}
\end{figure}

The spectrum of the radiation emitted by a rotating superluminal source is oscillatory with oscillations whose spacing increase with frequency~\cite{r1,r9}.  While the Bessel function describing synchrotron radiation has an argument smaller than its order and so decays exponentially with increasing frequency, the Bessel function encountered in~\cite{r9}, whose argument exceeds its order, is an oscillatory function of frequency with an amplitude that decays only algebraically.  Figure~\ref{fig7} shows that the spacing of the emission bands in the spectrum of the Crab pulsar fit the predicted oscillations for an appropriate choice of the single parameter $\Omega/\omega$.  The value of this parameter, thus implied by the data of~\cite{r15}, places the last peak of the oscillating spectrum at a frequency $(\sim\Omega^3/\omega^2)$ that agrees with the position of the ultraviolet peak in the spectrum of the Crab pulsar.  By inferring the remaining adjustable parameter $m$ in Eq.\ (\ref{eq:3}) from the observational data and by mildly restricting certain local properties of the source density ${\bf s}$, we are thus able to account for the continuum spectrum of the Crab pulsar over 16 orders of magnitude of frequency (Fig.~\ref{fig8}).

At radiation frequencies higher than $\sim10^{18}$ Hz, the Earth falls within the Fresnel zone of the Crab pulsar.  For an observer in this zone, the emission arises from a narrower radial extent of the source, and so has a higher degree of mean polarization, the higher the frequency.  The degree of circular polarization of such a high-frequency emission decreases with increasing frequency $\nu$ as $\nu^{-1/3}$, so that this emission is essentially 100 per cent linearly polarized at all phases, including off-pulse phases (see Eq.\ (15) of \cite{r1}). Note that, in this model, the intensity and polarization of the off-pulse emission, too, reflect the distributions of density and orientation of the emitting current around the cylinder $r=c/(\omega\sin\theta_P)$ within the pulsar magnetosphere~\cite{r11}.

Spectral differences of the different parts of the source that contribute to the emission detected at different rotation phases can result in the frequency dependence of the position angle of the high-frequency radiation.  Thus, one can have a linearly polarized high-frequency emission whose position angle is independent of phase but dependent on frequency (as observed in the Crab pulsar~\cite{r17}): independent of phase because the emitting current density can have a single non-zero component (e.g.\ the component $j_z$ parallel to the roation axis) at all points around the cylinder $r=c/(\omega\sin\theta_P)$, but frequency dependent because the distribution of the source as a function of ${\hat\varphi}$ can depend on frequency~\cite{r1}. 
   
\section{Comparison with observations}
\label{sec:5}

Characteristic features of the radiation generated by the superluminal emission mechanism described above suggest
\begin{enumerate}
\item[(i)]  that the extreme values of the brightness temperature ($\sim10^{37}$ K), temporal width ($\sim1$ ns), and source dimension ($\sim1$ m) of the giant pulses received from the Crab pulsar arise from the nonspherical decay of the intensity of this radiation with distancce,
\item[(ii)] that the observed micro- and nanostructures of the radiation beams received from the Crab pulsar arise from an incoherent superposition of sets of coherent, nondiffracting subbeams,
\item[(iii)] that the breadth of the spectrum of the radiation received from the Crab pulsar, and its proportionately spaced emission bands, reflect the algebraic (instead of exponential) rate of decay and oscillations of the superluminal counterpart of synchrotron spectrum, 
\item[(iv)] that concurrent `orthogonal' polarization modes with swinging position angles arise from the multiple images of a compact source and so should be associated with individual giant pulses,
\item[(v)] that the unpulsed part of the emission from the Crab is also generated by the pulsar itself, and
\item[(vi)] that the high degree of polarization of the high frequency radiation from the Crab pulsar reflects the decrease in size of the contributing part of the source with increasing frequency and a corresponding increase in degrees both of mean polarization and of linear polarization with increasing frequency. 
\end{enumerate}

\end{document}